\documentclass{aa}
\usepackage[round]{natbib}
\usepackage{graphicx}

\begin{document}

\title{Physical parameters of a relativistic jet at very high redshift:\\ the case of the blazar J1430+4204}
\author{P\'eter Veres \inst{1,2}, S\'andor Frey \inst{3,5}, Zsolt Paragi \inst{4,5}, Leonid I. Gurvits \inst{4}}
\authorrunning{Veres et al.}
\titlerunning{Physical parameters of J1430+4204}
\institute{Dept. of Physics of Complex Systems, E\"otv\"os University, P\'azm\'any P. s. 1/A, H-1117 Budapest, Hungary \\ 
\email{veresp@elte.hu} 
\and
Dept. of Physics, Bolyai Military University, POB 15, Budapest, H-1581, Hungary \and
F\"OMI Satellite Geodetic Observatory, POB 585, H-1592 Budapest, Hungary \and
Joint Institute for VLBI in Europe, POB 2, 7990 AA Dwingeloo, The Netherlands \and 
MTA Research Group for Physical Geodesy and Geodynamics, POB 91, H-1521 Budapest, Hungary}

\date{Received May 10, 2010; accepted Xxx XX, 2010}

\abstract
{The high-redshift ($z=4.72$) blazar J1430+4204 produced a major radio
outburst in 2005. Such outbursts are usually associated with the emergence of a
new component in the inner radio jet.}
{We searched for possible changes in the radio structure on milli-arcsecond
angular scales, to determine physical parameters that characterise the
relativistic jet ejected from the centre of this source.}
{We analysed 15-GHz radio interferometric images obtained with the Very Long
Baseline Array (VLBA) before and after the peak of the outburst.}
{We did not identify any significant new jet component over a period of 569
days. We estimated the Doppler factor, the Lorentz factor, and the apparent
transverse speed of a putative jet component using three different methods.
The likely small jet angle to the line of sight and our values of the apparent
transverse speed are consistent with not detecting a new jet feature.}
{}

\keywords{radio continuum: galaxies --- galaxies: active --- galaxies: jets --- quasars: individual (J1430+4204) --- techniques: interferometric}

\maketitle

\section{Introduction}

Active galactic nuclei (AGNs) are thought to harbour supermassive (up to
$\sim10^{9.5}M_{\odot}$) black holes. Accretion onto these black holes is
responsible for the extreme luminosity of AGNs over the whole electromagnetic
spectrum. Part of the infalling matter may be transformed into jets ejected
with relativistic speeds. The radio emission in radio-loud AGNs originates in 
these jets via synchrotron process. The appearance of the source strongly depends
on the orientation of the jet axis with respect to viewing direction.
If a jet points close to the line of sight
toward the observer, its brightness is significantly enhanced. For a review of
the unified schemes for radio-loud AGNs, see \cite{1995PASP..107..803U}.

A particular class of AGNs are blazars. They show large and rapid brightness variations,
from the radio to the gamma-ray regime. Among them, the BL Lac objects
have no strong emission lines characteristic of other AGNs in the optical spectrum. 
The radio polarisation of the blazars could be high and variable. According to the 
models of AGNs \citep{1995PASP..107..803U}, we look at these objects from almost
exactly the direction of their relativistic jets. 

J1430+4204 (B1428+4217) is a blazar with a flat radio spectrum, at an extremely
high redshift, $z=4.72$ \citep{1998MNRAS.294L...7H}. Long-term radio flux density
monitoring at 15~GHz revealed a significant brightening of J1430+4204, starting
in 2004 and reaching its peak in 2005 \citep{2006MNRAS.368..844W}. The object
increased its flux density by a factor of $3$ in about $4$ months (in the source
rest frame). At the same time, a corresponding outburst was not detected in X-rays 
with the XMM-Newton satellite, although a long-term X-ray spectral variability is 
evident in the source \citep{2006MNRAS.368..844W}. 
The X-ray spectrum of J1430+4204 suggests a large amount of intrinsic absorption 
with neutral hydrogen column density $N_{\rm H}\simeq10^{22}-10^{23}$~cm$^{-2}$
\citep{2004MNRAS.350L..67W,2006MNRAS.368..844W}.

The radio structure of J1430+4204 at the milli-arcsecond (mas) scale is
mainly compact as revealed by Very Long Baseline Interferometry (VLBI)
imaging observations at 5~GHz \citep{1999A&A...344...51P,2007ApJ...658..203H},
and 2.3 and 8.4~GHz (US Naval Observatory Radio Reference Frame Image Database,
USNO RRFID\footnote{\tt {http://rorf.usno.navy.mil/rrfid.shtml}}).  A weak
extension to the bright compact core is also seen in the W-SW direction.  On
scales three orders of magnitude larger, the central emission region is
dominant as well, but there is also a jet component at a $\sim$4$\arcsec$
angular distance from the brightness peak, in the W-NW direction \cite[Fig.~2
of][]{2008ASPC..386..462C}.  The pc- and kpc-scale structures are therefore
misaligned by $\sim$50$\degr$.  Imaging with the Chandra satellite revealed
X-ray emission also at the position of the kpc-scale radio jet component.  This
source has currently the highest redshift kpc-scale radio and X-ray jet known
\citep{2008ASPC..386..462C}. 

Total flux density outbursts are usually followed by an emergence of a new jet
component in the VLBI images of radio AGNs \cite[e.g.][]{2002A&A...394..851S}.
By observing J1430+4204 after the brightening, we aimed at detecting a new
component in a hope to have a zero-epoch point for a later measurement of its
apparent proper motion.  Due to the uniquely high redshift, the source
J1430+4204 would provide an important observational information for studies of
jet kinematics otherwise supported by the best-observed sample in the 15-GHz
MOJAVE (Monitoring of Jets in Active Galactic Nuclei with VLBA Experiments)
survey \citep{2009AJ....138.1874L} restricted mainly to $z<2.5$.  Reliable jet
proper motion measurements could not yet be made at redshifts as high as
$z$$\sim$5.  Information on the jet kinematics of J1430+4204 and other
high-redshift sources would eventually be an important contribution for
supplementing the apparent proper motion--redshift statistics as a cosmological
test \cite[e.g.][]{1988ApJ...329....1C,1994ApJ...430..467V,2004ApJ...609..539K}
with measurements at the extreme tail of the redshift distribution.

\begin{figure*}
\centering
\includegraphics[width=150mm]{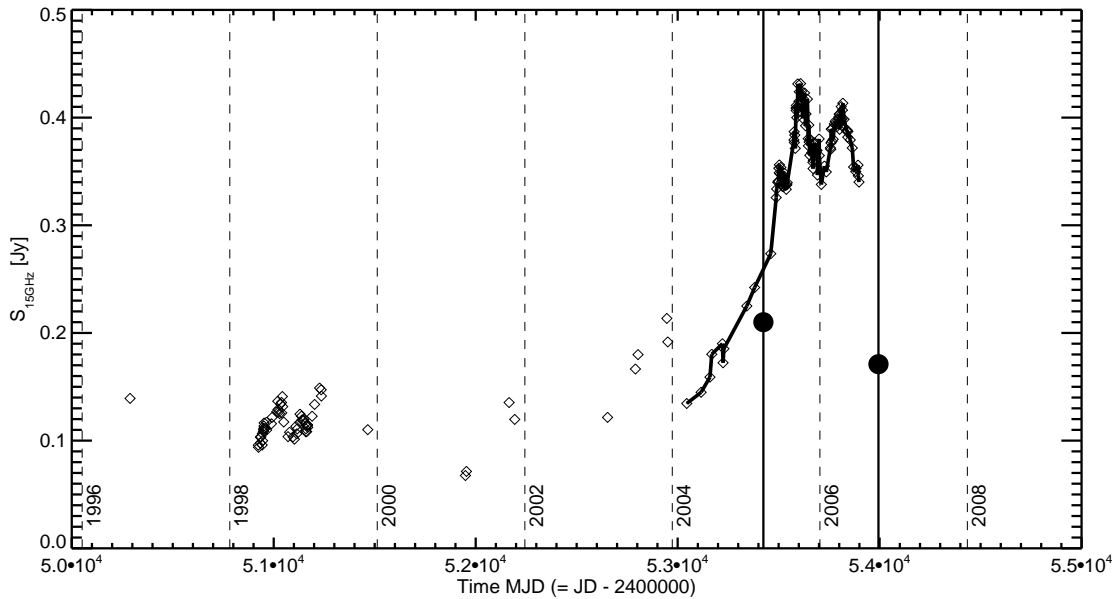}
\caption{The 15-GHz total flux density vs. time for J1430+4204 from the Ryle
Telescope monitoring (G. Pooley, priv. comm.). The points are connected in the
flaring phase for clarity. Solid vertical lines mark the times of the two VLBA
observations. The filled circles correspond to the measured VLBI flux
densities. Dashed lines mark calendar years as indicated. The typical flux
density error bars are smaller or comparable in size with the symbols.}
\label{lightcurve} 
\end{figure*}

In this study, we report on our high-resolution radio interferometric
observations of J1430+4204 (Sect.~\ref{observations}). We analyse the mas-scale
radio structure of the source, and determine the Doppler boosting factor and
the bulk Lorentz factor of the jet. We estimate the apparent proper motion of a
possibly emerging new jet component and conclude that it is below the detection
limit in our observations (Sect.~\ref{results}).  Finally, we put J1430+4204 in
context by comparing it with other similar AGNs at lower redshifts
(Sect.~\ref{discussion}).

\section{Observations and data analysis}
\label{observations}

We observed J1430+4204 at 15~GHz with the ten 25-m diameter radio telescopes of
the U.S. National Radio Astronomy Observatory (NRAO) Very Long Baseline Array
(VLBA) on 2006 September 15. The total observing time was 8~h. The data were
recorded in full circular polarisation with a bandwidth of 32~MHz. We also
obtained data taken earlier in a similar 15-GHz VLBA experiment (code BY019)
from the NRAO data archive. These observations were done on 2005 February 23,
prior to the peak of the total flux density curve (Fig.~\ref{lightcurve}) of
J1430+4204. The observing time was 5.8~h.  Both data sets were correlated at
the NRAO VLBA correlator in Socorro (New Mexico, USA).

We performed the standard VLBI calibration procedures for both experiments
using the NRAO Astronomical Image Processing System
\cite[AIPS;][]{1995ASPC...82..227D}. The visibility amplitudes were calibrated
using system temperatures and antenna gains measured at the antennas.
Instrumental delays were corrected using phase-calibration information provided
along with the data. Fringe-fitting was performed over 2-min solution
intervals.  Additional calibrator and fringe-finder sources (J0854+2006,
J0927+3902, J1256-0547, J1407+2827, J1426+3625 and J1642+3948) were also
observed. 

The calibrated visibility data were exported to and imaged with the Caltech
Difmap package \citep{1994BAAS...26..987S}. The conventional hybrid mapping
procedure with several iterations of CLEANing and phase (then amplitude)
self-calibration was applied. Our uniformly-weighted total intensity images of
J1430+4204 are displayed in Fig.~\ref{images}a (2005 February 23) and
Fig.~\ref{images}b (2006 September 15). 
Uniform weighting was applied to maximise the angular resolution. 
Naturally-weighted images did not reveal any other low surface brightness structure.
Identical coordinate scales and contour
levels are used in both images. The fractional linear polarisation of the VLBI
``core'' was $\sim$1\% and 2\% at the first and second epochs, respectively. We
detected no polarised emission from outside the core. 

The source total flux density at 15~GHz was monitored at the Ryle Telescope
\citep{1999MNRAS.308L...6F}. The light curve is shown in Fig.~\ref{lightcurve},
where the epochs of the two VLBA experiments, as well as the flux densities
measured with the VLBA are marked for comparison.

\section{Results}
\label{results}

\subsection{The compact radio structure of J1430+4204}

\begin{figure}
\centering
\includegraphics[width=62mm,angle=270,bb=50 80 520 770]{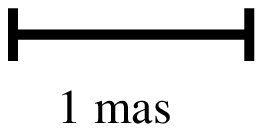}
\includegraphics[width=62mm,angle=270,bb=50 80 520 770]{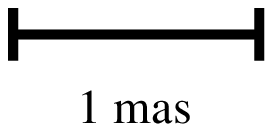}
\caption{15-GHz VLBA images of J1430+4204 on 2005 February 23 {\it ({\bf a},
top)} and 2006 September 15 {\it ({\bf b}, bottom)}. In the top image, the
peak brightness is 200 mJy/beam, the restoring beam is
1.17~mas~$\times$~0.52~mas at the position angle PA=$-3\fdg8$. In the bottom
image, the peak brightness is 159 mJy/beam, the restoring beam is
1.22~mas~$\times$~0.58~mas at PA=$-11\fdg95$. In both cases, the lowest contour
levels are at $\pm0.3$~mJy/beam, the positive contours increase by a factor of
2.} 
\label{images} 
\end{figure}

The mas-scale radio structure of J1430+4204 is dominated by a compact, nearly
unresolved core in the 15-GHz VLBA images, at both epochs (Fig.~\ref{images}).
Additionally, a diffuse emission region is seen with a low surface brightness,
out to $\sim$4~mas from the brightness peak, in the position angles between
about $-100\degr$ and $-150\degr$.  (Position angles are conventionally
measured from north through east.) This is most likely a steep-spectrum radio
emission from optically thin plasma along the jet. But we do not have spectral
information on the source at this spatial resolution, since our measurements
were taken at the single frequency of 15~GHz.  Notably, the position angle of
the faint emission region is consistent with the source extension observed with
VLBI at lower frequencies and lower resolution \cite[USNO
RRFID;][]{1999A&A...344...51P,2007ApJ...658..203H}.

Apart from the compact core, we did not detect any significant jet component of
which the proper motion could be reliably measured between the two observing
epochs separated by 569 days. 

We fitted elliptical Gaussian brightness distribution models to the radio core,
using the self-calibrated visibility data in Difmap. The resulting model
parameters are $S_1$=209~mJy flux density, $\vartheta_{1,1}$=0.141~mas major
axis, $\vartheta_{2,1}$=0.106~mas minor axis (full width at half maximum, FWHM)
at the first epoch, and $S_2$=166~mJy, $\vartheta_{1,2}$=0.143~mas,
$\vartheta_{2,2}$=0.059~mas at the second epoch. The fitted major axis position
angles ($-103\degr$ and $-101\degr$) indicate a W-SW extension of the innermost
radio structure, quite consistently with the weak mas-scale radio emission
(Fig.~\ref{images}).

The comparison of the total and VLBI flux densities  at the first epoch
(Fig.~\ref{lightcurve}) indicates that $\sim$50~mJy is completely resolved
out by the VLBA. Although total flux density monitoring data were not
available, by extrapolating the descending light curve, it is reasonable to
assume a similar difference at the second VLBA epoch. 

\subsection{Determination of the source parameters}

We used three methods for determining the following parameters of J1430+4204: the
apparent transverse speed of a possible blob in the jet 
\begin{equation}
\beta_{\rm app}= \frac{\beta \sin
\theta}{1-\beta \cos \theta},
\end{equation}
the Doppler boosting factor   
\begin{equation}  
\delta  = \frac{1}{\Gamma(1-\beta \cos \theta)},
\end{equation}
and the bulk Lorentz factor
 \begin{equation} 
\Gamma=\frac{\beta_{\rm app}^2 + \delta^2+1}{2 \delta} = (1-\beta^2)^{-\frac{1}{2}}.
 \end{equation} 
Here $\beta<1$ is the bulk speed of the material in the jet, expressed in the
units of the speed of light $c$. 

The most important physical parameter characterizing the jet flow is the bulk
Lorentz factor ($\Gamma$). We are also interested in the apparent tangential
velocity of the putative blob in the jet which we translate into proper motion.
The distance scale at this redshift is $6.497$ pc mas$^{-1}$. We assume a flat
cosmological model with $H_{\rm 0} = 72$~km~s$^{-1}$~Mpc$^{-1}$,
$\Omega_{\Lambda}=0.73$ and $\Omega_{\rm m}=0.27 $ throughout this paper.

Usually the Lorentz factor and the jet angle to the line of sight ($\theta$)
are determined from the measured Doppler factor, and the proper motion of jet
component(s) inferred from multi-epoch VLBI monitoring of radio AGNs. Since the
latter measurement is not available in our case, we took a different approach.
For the jet angle to the line of sight, $\theta=\arctan \left(\frac{2
\beta_{\rm app}}{\beta_{\rm app}^2 + \delta^2-1}\right)$, we assumed $3\degr$
as found from bulk Comptonization modelling of the observed X-ray spectrum of
J1430+4204 \citep{2007MNRAS.375..417C}.  We determined the Doppler factor using
three different methods, calculated the corresponding Lorentz factor, and
estimated the amount of possible jet component proper motion. The
inferred proper motion is indeed consistent with our non-detection. This in
turn supports the initial assumption that $\theta$ has a small value and
therefore this blazar is viewed nearly ``pole-on''.

\subsubsection{Parameters from the radio variability}

\begin{figure}
\centering
\includegraphics[width=85mm]{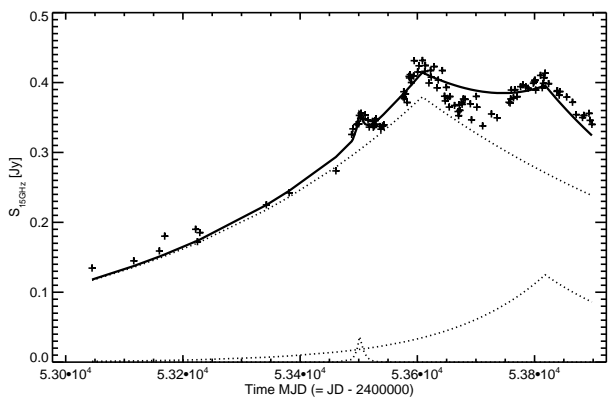}
\caption{The rising phase of the flare in the 15-GHz total flux density light
curve (crosses) of J1430+4204, 
fitted with a {three}-component exponential model (solid curve). 
{ Dotted curves show the individual components.}}
\label{fit} 
\end{figure}

\citet{2009A&A...494..527H} describe a method to determine the Doppler factor 
($\delta_{\rm var}$) from the total flux density variability of the source. 
Following this method, we fitted an exponential function to the brightest flare 
in the flux density curve (Fig.~\ref{lightcurve}) as
\begin{equation}
\Delta S(t)= \left\{
\begin{array}{ll}
\Delta S_{\rm max} e^{(t-t_{\rm max})/\tau},\quad t<t_{\rm max} \\
\Delta S_{\rm max} e^{(t-t_{\rm max})/1.3\tau},\quad  t>t_{\rm max}
\end{array}\right.
\end{equation}
where $t$ is the time, $t_{\rm max}$ is the epoch of the maximum flux density,
$\Delta S_{\rm max}$ is the maximum amplitude (Jy), and $\tau$ is the rise time
of the flare.  For the major flare occurred in 2005, we obtained $\Delta S_{\rm
max}$=0.380 ~Jy and $\tau$=$478\fd4$ from fitting a three-component model to the
light curve (Fig.~\ref{fit}).

We then determined the variability brightness temperature 
\begin{equation}
 T_{\rm b,var} = 1.548 \times 10^{-32} \frac{\Delta S_{\rm max} d_L^2}{\nu^2 \tau^2 (1+z)}
\end{equation}
where $d_L^2$ is the luminosity distance of the source (m) and $\nu$ is the observing frequency (GHz). 

For calculating the Doppler factor from variability 
\begin{equation}\label{dvar}
\delta_{\rm var} = \left({\frac{T_{\rm b,var}}{T_{\rm b,int}}}\right)^{\frac{1}{3}},\\
\end{equation}
we assume that energy equipartition holds between 
the particles and the magnetic field in the radio-emitting region 
\citep{1994ApJ...426...51R}. In this case the intrinsic brightness temperature is 
$T_{\rm b,int}=5\times 10^{10}$~K.

We obtained the variability brightness temperature $T_{\rm b,var} = {3.6}\times
10^{13}$~K, and the Doppler factor $\delta_{\rm var} = { 9.0}$.  The implied
apparent transverse speed is $\beta_{\rm app,var} = { 2.2}$ and thus the Lorentz
factor is $\Gamma_{\rm var} = {4.8}$. 

The apparent proper motion of a possible new jet component is $\mu =
0.02$~mas~yr$^{-1}$.  This means that during the time between the two VLBA
observations ($\Delta t = 569^{\rm d}$) the blob could have moved by
$\simeq$0.03~mas. This angular displacement is just below the limit which we
can possibly detect with the VLBA at this wavelength, about 10\% of the
restoring beam size \cite[cf.][]{2009AJ....138.1874L}.  Such a detection is
further complicated by the fact that the absolute astrometric position of the
core in our VLBA images is not known. Should the flare result in a slight
displacement of the source brightness distribution peak (well within the
restoring beam), the change remains unnoticed because self-calibration
effectively shifts the brightness peak of the VLBA image into the phase centre,
to the a priori source position that was assumed for the correlation of the
interferometric measurements.  Accurate relative astrometric registration of
the brightness peak could have been possible via phase-referencing observations
to an external radio source.  Alternatively, a well-defined stationary feature
further along the jet could have served as a positional reference, but there
seems no suitable component in our images (Fig.~\ref{images}).

\subsubsection{Parameters from the brightness temperature measured with VLBI}

The brightness temperature of a component in the radio source can be measured from
our VLBI data, using the following formula \cite[e.g.][]{1982ApJ...252..102C}:
\begin{equation}
T_{\rm b,vlbi} = 1.22\times 10^{12} (1+z) \frac{S}{\vartheta_1 \vartheta_2 \nu^2},
\end{equation}
where $S$ is the flux density (Jy), $\vartheta_{1}$ and $\vartheta_{2}$ are the major and minor
axes of the fitted Gaussian model component (FWHM, mas).  The $\nu$ observing
frequency is expressed in GHz. 

We again assume that the intrinsic brightness temperature is the
equipartition brightness temperature \citep{1994ApJ...426...51R,1999ApJ...511..112L}, 
$T_{\rm b,int} = T_{\rm eq} \simeq 5\times 10^{10}$ K. We will call the Doppler boosting factor derived
this way as the VLBI Doppler factor:
\begin{equation}\label{deq}
\delta_{\rm vlbi} = \frac{T_{\rm b,vlbi}}{T_{\rm b,int}}.
\end{equation}

This is valid if the brightness temperature is measured at the turnover
frequency. In our case, the turnover frequency in the broad-band radio spectrum
of is J1430+4204 is around 15~GHz \citep{2006MNRAS.368..844W}, close to our VLBA
observing frequency.

Our fitted elliptical Gaussian brightness distribution models to the core
suggest $T_{\rm b,vlbi,1} \simeq 4.3\times 10^{11}$~K  and $T_{\rm b,vlbi,2}
\simeq 6.0\times 10^{11}$~K at the two VLBA epochs. These values correspond to
 Lorentz factors of $\Gamma_1=4.6$ and $\Gamma_2=6.8$ for the two epochs.
Consistently with the first method based on variability, this one also predicts
a proper motion of $\mu = 0.02 - 0.03 $~mas~yr$^{-1}$.

\subsubsection{Estimating the intrinsic brightness temperature}

So far we assumed that the intrinsic brightness temperature of the source was equal 
to the equipartition value, $T_{\rm eq}=5\times 10^{10}$~K \citep{1994ApJ...426...51R}.
However, with two independent measurements of the apparent brightness temperature 
(from the variability and from the VLBI structure), and with the assumption that
the underlying Doppler factors are the same, we can calculate the intrinsic 
brightness temperature \citep{1999ApJ...511..112L}. Using Eqs.~\ref{dvar} and \ref{deq}, 
\begin{equation}
T_{\rm b, int} = (T_{\rm b,vlbi})^{\frac{3}{2}}  (T_{\rm b,var})^{-\frac{1}{2}}.
\end{equation}
At our two VLBI epochs, we get $T_{\rm b,int,1} = 4.7\times 10^{10}$~K and 
$T_{\rm b,int,2} = 7.7\times 10^{10}$~K, respectively.
Given the uncertainties in the estimates, these values are in very good agreement 
with the equipartition value.

\subsubsection{Parameters from the inverse Compton process}

By assuming that the X-ray radiation of the source results from inverse-Compton
scattered photons of the radio synchrotron radiation, one can estimate the 
Doppler factor \citep{1993ApJ...407...65G,1996ApJ...461..600G} as 
\begin{equation}
\delta_{\rm IC} = f(\alpha) (1+z) S_{\rm m} \times\left[\frac{\ln(\nu_{\rm
b}/\nu_{\rm op}) \nu_{\rm x}^{\alpha}}{S_{\rm X} \theta_{\rm
d}^{6-4\alpha}\nu_{\rm op}^{5-3\alpha}}\right]^{1/(4-2\alpha)} 
\end{equation}
where $f(\alpha) = -0.08 \alpha + 0.14$, $S_{\rm m}$ is the radio flux density
at the turnover frequency extrapolated from the optically thin part of the
spectrum, $\nu_{\rm b}$ is the synchrotron high-frequency cutoff assumed to be
$10^5$~GHz. The observed frequency of the radio peak is $\nu_{\rm
op}=15.3$~GHz. To apply this method, we also need X-ray data of the source
which we took from \citet{2007MNRAS.375..417C}.  The X-ray flux density is
$S_{\rm X}=2.32\times 10^{-7}$~Jy measured at $\nu_{\rm X}=1.25$~keV,
$\alpha=-0.38$ is the optically thin spectral index where the $S_\nu \propto
\nu^{\alpha}$ convention is used, and $\theta_{\rm d}$ is the angular diameter
of the source in mas. The fit for radio peak frequency and the extrapolation from the
optically thin part was carried out using the data by \citet{2006MNRAS.368..844W}. 

Here we also need to assume that the VLBI observations were carried out at the
turnover frequency, which holds for our case. We find the value of the
inverse-Compton Doppler factor to be $\delta_{\rm IC,1}\simeq 11$ and
$\delta_{\rm IC,2}\simeq 15.9$ for the first and second VLBA epochs,
respectively.  Assuming again $\theta = 3^\circ$ \citep{2007MNRAS.375..417C},
the Lorentz factor and the apparent transverse speed for the two epochs are:
$\Gamma_{\rm IC,1}\simeq 6.1$ and $\Gamma_{\rm IC,2}\simeq 10.3$; $\beta_{\rm
app,1}=3.4$ and $\beta_{\rm app,2}=8.6$. At the distance of the source, this
translates to an apparent proper motion of $\mu_1 \simeq 0.03$~mas~yr$^{-1}$
and $\mu_2 \simeq  0.07$~mas~yr$^{-1}$.  In 569 days, the putative jet
component would have moved by $\simeq 0.04 - 0.1$~mas.


\section{Discussion}
\label{discussion}

We estimated the Doppler factor in the blazar J1430+4204 using different methods 
based on radio variability ($\delta_{\rm var}$), VLBI imaging assuming equipartition ($\delta_{\rm vlbi}$), 
and the inverse Compton process ($\delta_{\rm IC}$). These estimates are summarised in
Table~\ref{parameters}, along with the other source parameters derived.

\begin{table}
\caption[]{Parameters of J1430+4204 estimated from radio variability, VLBI imaging and the inverse Compton process.}
\label{parameters}
\centering 
\begin{tabular}{lcccc}        
\hline\hline                 
Method  & $\delta$ & $\Gamma$ & $\beta_{\rm app}$ & $\mu$ (mas~yr$^{-1}$) \\
\hline                       
var     &  ${ 9.0}$		& ${ 4.8}$  		&   ${ 2.2}$ 		& $0.02$ \\
VLBI    &  $8.6$-$12.0$ 	& $4.6$-$6.8 $  	&   $2.0$-$4.2$ 	& $0.02$-$0.03 $ \\
IC      &  $11.0$-$15.9$ 	& $6.1$-$10.3$  	&   $3.4$-$8.6$ 	& $0.03$-$0.07$ \\
\hline   
\end{tabular}
\end{table}

While the mean $\delta_{\rm vlbi}/\delta_{\rm IC}$ ratio is in the order of unity
for the AGN sample of \citet{1996ApJ...461..600G}, there are differences up to
a factor of $\sim$4 in the Doppler boosting factors derived by the two
independent methods. Errors come from the uncertainties of the input
parameters, and larger inconsistencies may reflect deviations from the
equipartition in certain sources.  In the case of J1430+4204, our three
estimates of the Doppler factor are remarkably similar, the values agree within
a factor of $\sim$2.

To estimate the bulk speed of the jet material ($\beta$) and the corresponding
Lorentz factor ($\Gamma$), we assumed $\theta=3\degr$ for the jet angle to the
line of sight, found from bulk Comptonisation modelling of the X-ray spectrum by
\citet{2007MNRAS.375..417C}. At a given Doppler factor, $\sin\theta \leq 1/\delta$
\citep[e.g.][]{1995PASP..107..803U}, which in our case constrains the allowed range of
the viewing angles to  $\theta \la 6 \degr$. The value of $\Gamma$ is quite
insensitive to the actual $\theta$ in most of the possible range of viewing
angles. Therefore our conclusions remain valid even if we allow for somewhat
different values of $\theta$.

From measuring the source variability and from imaging the mas-scale radio structure 
at nearly the same time, we could determine the intrinsic brightness temperature of 
J1430+4204. Our values ($T_{\rm b,int,1} = 4.3\times 10^{10}$~K and 
$T_{\rm b,int,2} = 7.1\times 10^{10}$~K at two different epochs)
are consistent with those obtained for a larger sample of lower-redshift AGNs 
\citep{1999ApJ...511..112L}, suggesting that $T_{\rm b,int} \la 10^{11}$~K
in compact extragalactic radio sources.

How is the other intrinsic parameter, the Lorentz factor of the jet in
J1430+4204 ($\Gamma \simeq 5-10$) compared to those values measured in other
AGNs? The systematic multi-epoch 15-GHz VLBA monitoring of a large
flux-density-limited sample of radio-loud AGNs in the 2-cm VLBA Survey and the
MOJAVE survey \citep{2004ApJ...609..539K, 2007ApJ...658..232C,
2009AJ....138.1874L} revealed that Lorentz factors can be as high as
$\Gamma\sim50$, but the sources typically have values between 5 and 10.  The
overwhelming majority of the MOJAVE sources are at $z<2.5$; the most distant
object in the sample is J0646+4451 at $z=3.4$. As far as the Lorentz factor is
concerned, our high-redshift object is therefore similar to those studied in
the MOJAVE survey.

The maximum apparent speed vs. redshift plot for the MOJAVE quasars
\cite[Fig.~8 of][]{2009AJ....138.1874L} shows that most of the values are found
in the range $2<\beta_{\rm app}<30$.  The estimated apparent speed of
J1430+4204 fits well in this picture and is closer to the low values. 

The estimate for the jet apparent proper motion ($\mu\simeq0.03$~mas~yr$^{-1}$)
obtained from our analysis for J1430+4204 is small compared to other sources at
lower redshifts.  This result fits the observed angular velocity--redshift
relation and is consistent with the cosmological interpretation of the
redshifts \citep{2004ApJ...609..539K}.

\section{Conclusion}
\label{conclusion}

The high-redshift ($z=4.72$) blazar J1430+4204 produced a { major} radio
flux density outburst in 2005 (Fig.~\ref{lightcurve}). About a year after the
peak of the 15-GHz flux density curve, we imaged the source with the VLBA at
this frequency.  We also analysed similar archive VLBA data taken during the
rise of radio light curve.  At both epochs separated by 569 days, the mas-scale
radio structure of the source was remarkably similar: a compact optically thick
``core'' and a weak extension to W-SW (Fig.~\ref{images}). The core could be
fitted with elliptical Gaussian model components.

We searched for a new emerging component usually observable after large radio
outbursts in AGNs. Based on our VLBA imaging, we did not detect any new jet
component to be associated with the event.  Assuming a small jet angle to the
line of sight, we used three different methods to estimate the expected proper
motion of such a component. These gave consistently small values for the proper
motion.  We found that our time base and angular resolution were insufficient
to distinguish any new blob in the jet. { Supposing that the jet was
launched in 2004, the elapsed time until the second VLBA observation is $950$
days. During this time the putative blob should have moved by 0.05 mas.}

Our estimates for the intrinsic physical parameter describing the jet, the bulk
Lorentz factor ($\Gamma \simeq 5-10$) are comparable to the typical values
found for other well-studied samples of blazars
\cite[e.g.][]{2009A&A...494..527H,2009AJ....138.1874L}.  
The intrinsic brightness temperature is close to the equipartition value
and is fully consistent with brightness temperatures measured in 
other sources \citep[$T_{\rm b,int} \la 10^{11}$~K;][]{1999ApJ...511..112L}

According to our estimates of the apparent proper motion, the emergence of a
new jet component associated with the { major} 2005 outburst in J1430+4204 could
possibly be detected with the angular resolution of ground-based VLBI at 15~GHz
over a time span of $\sim$10 years.

\begin{acknowledgements} 
The National Radio Astronomy Observatory (NRAO) is a facility
of the National Science Foundation operated under cooperative agreement by
Associated Universities, Inc.
We are grateful to Guy Pooley for sharing the Ryle Telescope data with us. 
{ We thank the referee, Esko Valtaoja for his comments.}
This work was supported by the Hungarian Scientific Research Fund (OTKA grants
K077795 and K72515).  This research has made use of the United States Naval
Observatory (USNO) Radio Reference Frame Image Database (RRFID), and the
NASA/IPAC Extragalactic Database (NED) which is operated by the Jet Propulsion
Laboratory, California Institute of Technology, under contract with the
National Aeronautics and Space Administration.
\end{acknowledgements}

\bibliographystyle{aa} 
\bibliography{j1430}

\end{document}